# Dust charging in dynamic ion wakes

Lorin Swint Matthews, Dustin L. Sanford, Evdokiya Kostadinova, Khandaker Sharmin Ashrafi, Evelyn Guay, Truell W. Hyde

Center for Astrophysics, Space Physics, and Engineering Research, Baylor University
One Bear Place 97316, Waco, Texas 76798-7316

Abstract

Micron-sized dust grains have been successfully employed as non-perturbative probes to measure variations in plasma conditions on small spatial scales, such as those found in plasma sheaths. The dynamics of the grains can be used to map the forces due to electric fields present in the sheath, but the particle charge and electric field are difficult to measure independently. The problem is further complicated by the ion wake field which develops downstream of the dust grains in a flowing plasma. Within a sheath, ions are accelerated towards the charged boundary, and this ion flow creates a positively-charged spatial region downstream of the dust grain, called the ion wake. The ion wake in turn modifies the interaction potential between the charged grains. Here we use a molecular dynamics simulation of ion flow past dust grains to investigate the interaction between the charged dust particles and ions. The charging and dynamics of the grains are coupled self-consistently and derived from the ion-dust interactions, allowing for detailed analysis of the wakefield-mediated interaction as the structural configuration of the dust grains changes. The decharging of a dust grain as it moves through the wake of an upstream particle and the attractive ion wakefield force are mapped for a range of ion flow speeds.

I. Introduction

Complex (or dusty) plasma systems are a special type of low-temperature plasmas, where micron-sized particles (or dust) can levitate and self-organize into strongly coupled fluids, field-aligned chains, and crystalline solids [1]–[5]. Such structures are commonly formed in streaming plasmas, where the ion and electron particles flow along the direction of an externally-applied electric field. When dust grains are immersed in such an environment, they become negatively charged, which makes them repel each other. However, the dynamical interaction of the dust particles with the ion flow often leads to attractive and non-reciprocal forces, which affect structure formation in those systems [6]–[8]. Therefore, the study of self-organization, dynamics, and stability of dusty plasmas requires a proper understanding of the dust interaction with a streaming ion flow.

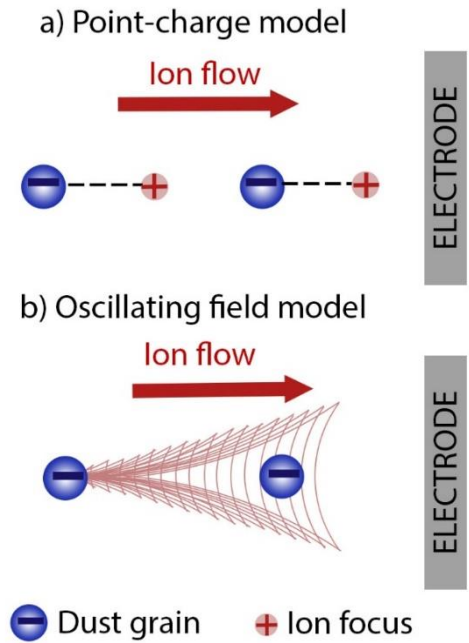

Fig. 1. Modeling the ion wakefield: a) a point-charge model, b) an oscillating ion field model.

As the ions pass near a negatively charged dust particle, depending on their velocity and proximity, their trajectories can be deflected, resulting in the formation of a wakefield structure downstream of the grain. A simplified theoretical approach for modeling this effect is to represent the ion wakefield as a point-like region of positive space charge (called the wakefield focus) located downstream of the grain (Fig. 1a) [9]–[11]. In this model, the wakefield focus can provide an attractive force for negatively charged dust particles located downstream, yielding field-aligned structure formation. Although the point-charge representation has been successful in demonstrating the non-reciprocal character of the grain-grain interaction, its application is often limited to strongly-coupled dusty plasma structures where the downstream position and magnitude of the point charge can be assumed constant. An alternative approach is the oscillating wakefield approximation [12], [13], where the downstream dust particle is assumed to interact with a superposition of ion acoustic waves generated by the upstream grain (Fig. 1b). Although this model is useful in the study of dusty plasma waves and instabilities, it cannot reveal the structure of the ion wakefield at the kinetic level. Since the point-charge mechanism emphasizes the electrostatic nature of the grain-grain interaction, while the oscillating wakefield representation highlights the wave-like character of the ion flow, combining features of the two models will enable proper mapping of the ion wakefield structure.

To accomplish this, advanced numerical simulations are needed to resolve the trajectories of the individual ions and electrons, as well as the dynamics of the interacting dust grains in dusty plasmas. Possible candidates include Monte Carlo [14] or Particle-In-Cell (PIC) [15]–[17] simulations, which are often used to determine the structure of the ion wakefield downstream of the dust grain and calculate the resulting nonlinear grain-grain interactions [18], [19]. The study of dust charging in plasma and the formation of the ion wakefield structure has also been advanced by first-principle PIC simulations modeling the trajectories of both ions and electrons [20]–[23].

The features of the ion wakefield behind charged dust grains have also been explored using molecular dynamics (MD) simulations of the ions within the plasma flow [24]–[26]. The efficiency of this numerical technique was recently enhanced by the development of high-performance GPUs, capable of parallel processing with more than 1000 processors. Piel [24] introduced a molecular asymmetric dynamics (MAD) code, where ions move in the "naked" Coulomb potential of dust grains, while interacting with each other through Yukawa force accounting for electron shielding. The MAD code employs super-ions with the same charge-to-mass ratio (and therefore dynamical behavior) as a single ion, which improves time efficiency and allows for large simulation volumes. Despite these approximations, the MAD technique can reproduce important features of the ion wake potential in reasonable agreement with previous PIC simulations.

The numerical techniques discussed so far commonly treat dust grains as static obstacles within the streaming plasma. However, investigation of phase transitions and fluid phenomena in dusty plasmas requires a model where the dust grains are both free to move and allow their charge to vary in the field of streaming ions. In this case, the ion wakes are modified by the presence of the downstream dust particles and the charging of the downstream grains is, in turn, affected by the location of the ion wakes. Therefore, the ion wake formation and grain charging are coupled and should be treated in a self-consistent manner. This paper introduces a molecular dynamics (MD) simulation which resolves the motion of both dust grains and ions and simultaneously allows for charging of the dust particles by collisions with ions in a flowing plasma.

The paper is organized as follows. The details of the numerical model are presented in Section II, including the treatment of the ions, the dynamics of the dust, and the calculation of dust charging. The use of the numerical model is illustrated by modeling the behavior of two vertically aligned dust grains confined within a glass box placed on the lower electrode of a GEC rf reference cell. The lower particle is perturbed using a laser, causing it to oscillate within the wake of the upper particle. The results of these simulations are presented in Section III. Section IV provides a summary and conclusions.

## II.     Numerical Model

The numerical model DRIAD (Dynamic Response of Ions And Dust) is a molecular dynamics simulation designed to resolve the motion of both the ions and the dust grains on their individual timescales, while allowing the dust charge to vary in response to the changing ion density in the ion wakefield. Following the method described by Piel [24], the forces between pairs of ions and among ions and dust particles are treated in an asymmetric manner. The ion-ion interactions are assumed to arise from a Yukawa-type potential with shielding provided by the electrons, while the ion-dust interactions are assumed to be Coulomb interactions. Comparisons with PIC simulations [27] show that this asymmetric treatment provides a reasonable agreement for the equilibrium potential distribution and interparticle forces.

Ions will typically reach an equilibrium distribution within one plasma period

$$\tau_i = 2\pi \sqrt{\epsilon_0 m_i / n_i e^2} \qquad (1)$$

where $m_i$ is the mass of an ion, $n_i$ is the number density of the ions, and $e$ is the elementary electric charge. For the range of plasma densities characteristic of the sheath of a rf discharge, the plasma period is typically $\tau_i \approx 1\ \mu s$. The DRIAD code resolves the motion of the streaming collisional ions on the ion timescale, typically $\Delta t_i = \tau_i/100$ s, with the simulation providing enough time steps to cover one ion plasma period. The ions are then frozen, and the dust is advanced one dust time step, $\Delta t_d = 10^{-4}$ s, with the appropriate parameters associated with the ions averaged over the elapsed ion time steps. The present model does not resolve the motion of individual electrons. Instead, for the purposes of calculating the ion-ion interaction force and dust charging, the electrons are assumed to have a Boltzmann distribution.

### A.     Treatment of Ions

Given the high number density of ions in the plasma, the numerical model computes the trajectories of superions, each of which represents a cloud of ions with the same charge-to-mass ratio (and hence equation of motion) as a single ion. Allowing ~100 ions per superion results in reasonable CPU time (typically 12-24 hours to resolve one second of dust motion). The equation of motion for an ion with charge $q_i$ is given by

$$m_i \vec{\ddot{r}} = -q_i \vec{E}_i + \vec{F}_{in} \qquad (2)$$

where the electric field $\vec{E}_i$ consists of contributions from the other ions in the simulation, the charged dust particles, and any electric fields present in the plasma (for example, the electric field in the sheath of a rf discharge) and $\vec{F}_{in}$ is the ion-neutral collision force.

The ion-ion interactions are calculated assuming a Yukawa potential

$$\Phi_Y(r_i) = \frac{1}{4\pi\epsilon_0} \sum_j \frac{q_j}{r_{ij}} \exp(-\frac{r_{ij}}{\lambda_{De}}), \qquad (3)$$

where $q_j$ is the charge on ion $j$, $r_{ij}$ is the distance between ion $i$ and ion $j$, and the Boltzmann electrons provide the screening with shielding length $\lambda_{De}^2 = \epsilon_0 k_B T_e / (n_e e^2)$. Here $k_B$ is the Boltzmann constant, $T_e$ is the temperature of the electrons, and $n_e$ is the number density of the electrons. As the region surrounding the negatively charged dust is depleted in electrons, the interaction between the ions and dust is calculated using a Coulomb potential

$$\Phi_C(r_i) = \frac{1}{4\pi\epsilon_0} \sum_d \frac{Q_d}{r_{id}}, \qquad (4)$$

where $r_{id}$ is the distance between the ion and dust grain with charge $Q_d$.

As we are particularly interested in the charging and dynamics of multiple dust particles within a flowing plasma, we define a cylindrical simulation region with the cylinder's axis aligned with the ion flow. Ions which leave the simulation region or are absorbed by a dust grain are reinjected on the boundary in a manner which is consistent with a shifted Maxwellian distribution with drift velocity $v_{dr}$ and number density $n_{i0}$. The algorithm to determine the velocity and position of the incoming ions is adapted from the algorithm presented in [28] for insertion of ions on a spherical boundary.

The ions in the cylindrical region are subject to a confinement force from the assumed infinite homogeneous distribution of ions outside the simulation region. In Piel [24] an analytic expression for the electric field inside of a spherical cavity in homogeneous Yukawa matter [29] was used to provide this boundary condition. A closed-form analytic expression does not exist for the electric field inside a cylindrical cavity. Instead, the electric field from these ions is determined by first numerically calculating the Yukawa potential of a homogenous distribution of ions of density $n_0$ within the cylindrical simulation region. This potential is then subtracted from a constant uniform background potential yielding the potential within a cylindrical cavity inside the homogeneous Yukawa material. The confining electric field is calculated from the negative gradient of this potential. This is done once at the beginning of the simulation on a sufficiently fine grid, and the confinement force acting on each ion at a given location is interpolated from the electric field calculated on the grid points.

At high pressures, ion–neutral collisions significantly reduce particle charge, [30]–[32]. Ion-neutral collisions also create a drag force which balances the acceleration due to external electric fields (such as the electric field in the sheath of an rf plasma or the constant electric field in a DC plasma), leading to a constant drift velocity $v_d$ in the direction of the applied field. In the present model, the ion-neutral collisions are calculated using the null-collision method [33]. The data for the Ar-Ar$^+$ collision cross sections are taken from the Phelps database (hosted by the LxCat project) [34], while the data for the Ne-Ne$^+$ cross sections are obtained from Jovanović et al. [35].

B. Dust Dynamics

The equation of motion for a dust grain with mass $m_d$ and charge $Q_d$ in a typical laboratory experiment is given by

$$m_d \ddot{x} = F_{dd} + F_{id} + m_d g + Q_d E - \beta \dot{x} + \zeta r(t). \tag{5}$$

Where $F_{dd}$ is the force between dust grains, $F_{id}$ is the force exerted by the ions on a dust grain, $m_d g$ is the gravitational force, $E$ is the confining electric field within the region, $\beta$ is the neutral drag coefficient, and $\zeta r(t)$ is a thermal bath, with $r(t)$ a random number uniformly distributed between -1 and 1. The force between pairs of dust grains $F_{dd}$ is assumed to be a Coulomb interaction, since screening is mainly provided by the ions. $F_{id}$ includes the forces from all ions in the simulation region calculated from the Coulomb interaction potential given in Eq. (4), averaged over the elapsed ion time steps, as well as the force from the homogeneously distributed ions outside the simulation region, as described above.

The confining force $Q_d E = Q_d(E_r \hat{r} + E_z \hat{z})$ arises from the electric fields within the simulation. In the present paper, we consider a laboratory experiment where the dust is levitated against the force of gravity $m_d g$ by the vertical electric field $E_z$ in the sheath above the lower electrode of a GEC rf reference cell. The sheath electric field is typically assumed to vary linearly with distance from the electrode. The electric field profiles obtained from a fluid model of the plasma within CASPER's GEC cell shows that this is a good approximation for the vertical region where the dust floats, and that the electric field steepens with increasing power [36]. Here we are interested in laboratory experiments with extended vertical dust structures, where confinement in the horizontal direction is provided by the charged walls of a glass box placed on the lower electrode, which has been shown to be very nearly radial near the middle of the box [37].

The damping factor $\beta$ depends on the neutral gas pressure and temperature, with

$$\beta = \delta \frac{4\pi}{3} a^2 n \frac{m_g}{m_d} \sqrt{8 k_B T_g / \pi m_g} \tag{6}$$

where $\delta = 1.44$ (measured for melamine formaldehyde dust in Argon), $a$ is the dust radius, $n$ the gas number density, $m_g$ the molecular mass of the gas, $T_g$ the gas temperature, and $m_d$ is the mass of the dust.

A thermal bath is realized by subjecting the particles to random kicks, with the maximum acceleration imparted by a kick

$$\zeta = \sqrt{\frac{2 \beta k_B T_g}{m_d \Delta t_d}} \tag{7}$$

where $\Delta t_d$ is the dust time step.

### C. Dust Charge

Typically dust grains charge through collisions with electrons and ions in the plasma, which constitute currents to the grain's surface that depend upon the dust surface potential. In low-density plasmas, where the ions can be treated as collisionless, the grain charge is often calculated using orbital motion limited (OML) theory [38]. However at higher pressures, charge-exchange collisions between ions and neutral gas particles can generate ions which are trapped around the dust particle, causing OML theory to over-predict the particle charge [30]–[32]. Here we use a combined MD and OML approach to determine the dust charge.

The electrons, which are not explicitly modeled, are assumed to have a Boltzmann distribution and the electron current from OML theory is used

$$I_e = 4\pi a^2 n_e e \left(\frac{kT_e}{2\pi m_e}\right)^{1/2} \exp\left(-\frac{e\Phi_d}{k_B T_e}\right). \tag{8}$$

where $n_e$, and $m_e$ are the electron density and mass, the grain radius is $a$, and $\Phi_d$ is the dust surface potential.

The charge accumulated during the time step $\Delta t_i$ due to the electron current is

$$\Delta Q_{de} = I_e \Delta t_i. \tag{9}$$

The positive charge accumulated is the number of ions which collide with the grain during the time step, $\Delta Q_{di} = N_{ic} q_i$, thus the total charge accumulated at each time step is $\Delta Q = \Delta Q_{de} + \Delta Q_{di}$. A time history of the dust particle charge at each ion time step, where $\Delta t_i = 10^{-9}$ s is shown in Fig. 2 for a 10.4 μm-diameter grain in neon plasma with electron temperature $T_e = 3.4$ eV, ion temperature $T_i = 0.01 T_e$, $n_e = n_i = 2 \times 10^{14}$ m$^{-3}$, and neutral gas pressure $P = 40$ Pa. For this case, the dust charge reaches equilibrium within one plasma ion period $\tau_i = 1.5$ μs. As expected, the dust charge is considerably reduced from that predicted by OML theory for a collisionless plasma ($2.5 \times 10^4$ e$^-$ for these conditions). The charge is averaged over the elapsed ion time steps to obtain $Q_{avg}$ before calculating the dust dynamics on the dust time step (shown in light blue). It is important to note that the stochastic nature of the charging can produce large fluctuations in the charge which occur on a timescale that is too fast for the dust to respond. Additionally, these fluctuations can persist over a period longer than the ion plasma period. In this case, the time averaged charge is $Q_{avg} = 7640$ e$^-$, with a standard deviation $\sigma = 320$ e$^-$.

Theoretically, the charge variance is given by [39]

$$\sigma^2 = \frac{4\pi\epsilon_0 a k_B T_e}{e^2}\left(1 - \left(1 + \frac{T_i}{T_e} - \frac{e\Phi_d}{k_B T_e}\right)^{-1}\right), \tag{10}$$

which predicts a much smaller standard deviation $\sigma = 69$ e$^-$ for these conditions, as compared to the variance calculated from the charge history. To smooth out these large fluctuations on the dust time step, which is nearly 100 times longer than $100\tau_i$, the dynamic dust charge is calculated from a weighted average of the dust charge at the previous dust time step and the average charge at the current dust time step

$$Q_d(t_d) = 0.95\, Q_d(t_d - 1) + 0.05\, Q_{avg}(t_d). \tag{11}$$

In Fig. 2, $Q_d$ is represented by the red dashed line. Although the dynamic dust charge lags behind the charge calculated on the ion time step, it reaches the equilibrium charge within about 50 dust time steps. After reaching equilibrium, the time averaged charge is $Q_d$ = 7560 e⁻, with a standard deviation $\sigma$ = 76 e⁻, which reasonably agrees with the theoretically predicted charge variance.

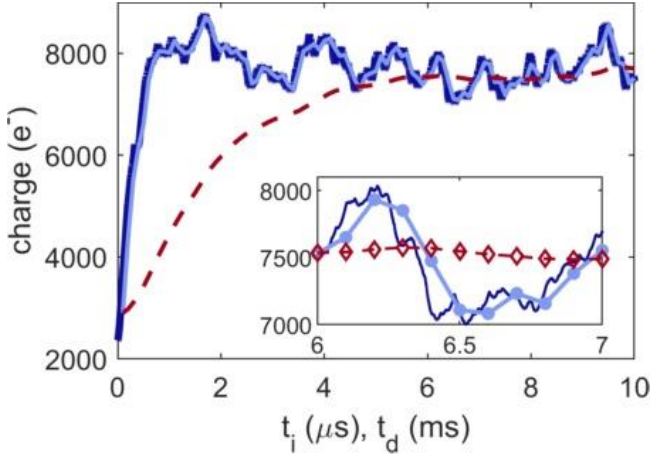

Fig. 2. Charge on a 10.4 $\mu$m dust grain as a function of time. The dark blue line shows the charge at each ion time step, the light blue line is the charge averaged over the last 100 ion time steps, and the red dashed line is the smoothed charge $Q_d$ as calculated in Eq. 11 for each dust time step. The inset shows a detail of the charge history with the markers indicating the averaged values. For these calculations the plasma conditions are assumed to be $T_e$ = 3.4 eV, $T_i = 0.01 T_e$, $n_e = n_i = 2 \times 10^{14}$ m⁻³, $P$ = 40 Pa, neon gas.

III. Results

In this section, we illustrate the connection between dust particle charging and the ion wakefield by modeling the interaction between two dust grains confined within the sheath above the lower electrode of a GEC reference cell. The simulation parameters aim to reproduce an experiment performed in the CASPER lab, where the horizontal confinement for the particles was provided by the charged walls of a glass box placed on the lower powered electrode. The system power and pressure were adjusted to increase the horizontal confinement, allowing the particles to arrange themselves into a vertical pair. A Coherent Verdi V-5 laser was used to perturb the lower particle, and the motion of the two particles was then recorded at 500 or 1200 frames per second (fps) using a side-mounted high-speed CCD (Photron) camera and a microscope lens. The experimental setup is shown in Fig. 3. In this experiment, analysis of the dust motion was used to determine the acceleration of the particles and to reconstruct a map of the electric field within the region [40]. However, calculation of the electric fields using this method assumes that the particle charges remain constant during the interaction.

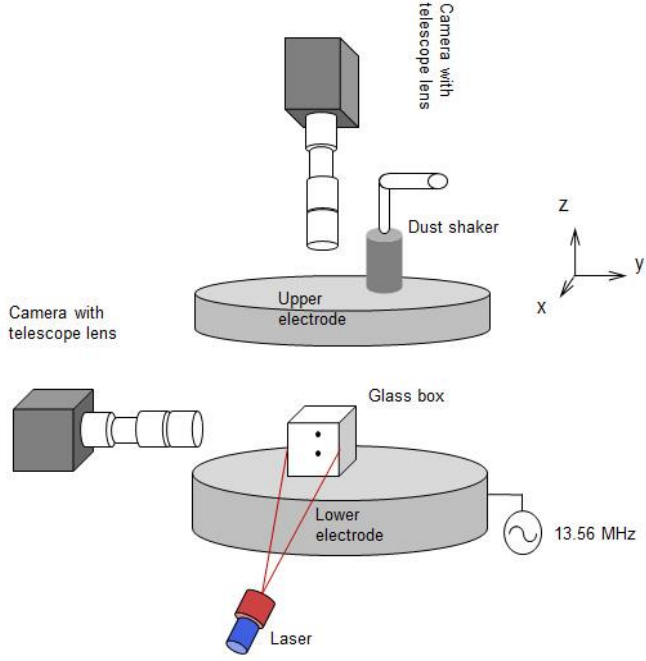

Fig. 3. Vertical dust pair experiment in a GEC rf reference cell. The upper electrode is grounded and the glass box sits on the lower powered electrode. The two vertical and horizontal cameras track the motion of the dust particles within the box.

To examine this experiment numerically, we model the charging and dynamics of the two-particle system for three different ion drift velocities, $v_{dr}$ = 0.4, 0.6, and 1.0 M, where the Mach number M is in units of the ion sound speed $c_s = \sqrt{k_B T_e/m_i}$. The simulation parameters assume argon gas at a pressure of 6.67 ± 0.10 Pa and a plasma bulk density $n_0 = 2 \times 10^{14}$ m$^{-3}$. The electron temperature is taken to be $T_e$ = 5 eV and the ion temperature is $T_i$ = 290 K. The electron Debye length for this set of parameters is $\lambda_{De}$ = 1.2 mm and the ion Debye length $\lambda_{Di}$ = 83 $\mu$m. The cylindrical simulation region has a radius of 1.5 $\lambda_{De}$ and height of 5 $\lambda_{De}$. Two dust particles are placed in a simulated glass box with the lower particle oscillating vertically within the wake of the upper particle. At time $t$ = 0.15 s the lower dust particle is given a horizontal acceleration of 0.5 m s$^{-2}$ for a time $\Delta t\_laser$ = 0.05 s to simulate a laser pushing the particle outside of the wake. Fig. 4 (Mulimedia view) shows the path of the lower particle, P2, relative to the upper particle, P1 for three separate runs at the three ion drift velocities. The progression of time is shown as the path changes from light to dark. In the latter two cases, P2 and P1 switch places briefly. For each condition, we examine the characteristics of the ion wake field as the particles interact, the charge as a function of the particle separation, and the acceleration caused by the ion-mediated dust-dust interaction. In the numerical model, the charge can be mapped during the interaction, allowing a more accurate electric field map to be constructed.

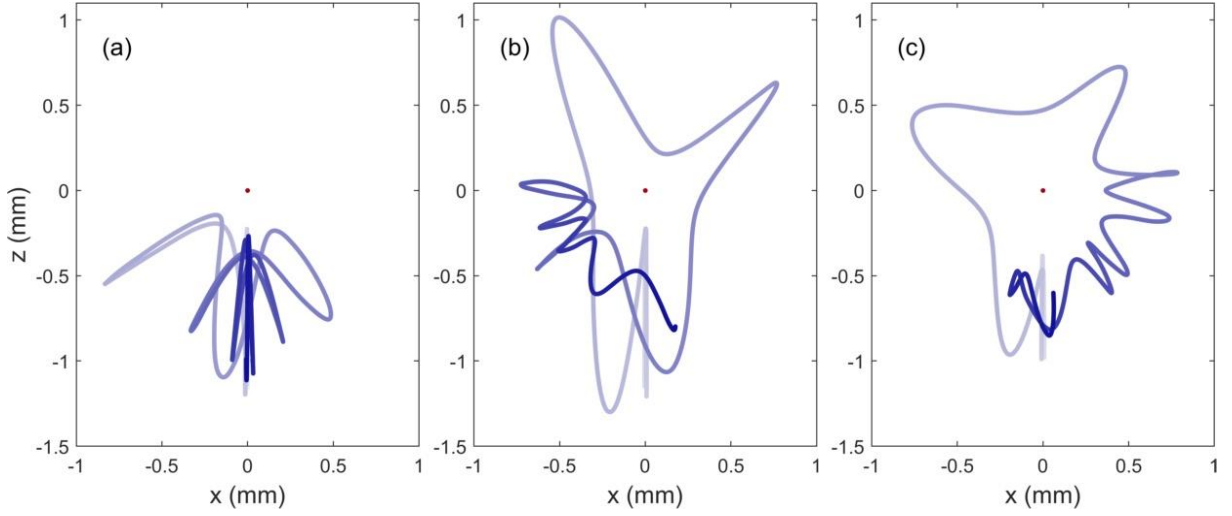

Fig. 4 (Multimedia view). Trajectory of second particle P2 relative to particle P1, marked by a red dot, for (a) $v_{dr}$ = 0.4 M, (b) $v_{dr}$ = 0.6 M, and (c) $v_{dr}$ = 1.0 M. Time progression is indicated by shade progressing from light to dark. (Multimedia view).

A.   Ion Wakefield

The ion density maps for the various drift velocities are shown in Fig. 5 (Multimedia view) for three different particle separations. At low ion drift velocities (left column), each dust particle has a distinct ion cloud surrounding it, except when the two particles are very close together, $\Delta z = 0.3\ \lambda_{De}$, as shown in the top row. As the ion velocity increases, the ion density cloud is elongated in the direction of ion flow, and the focusing region moves downward. Thus, for the highest ion velocities, only a single high-density cloud is formed beneath the lower particle.

Maps of the combined ion dust potential for the same conditions are shown in Fig. 6 (Multimedia view). The total potential at each point $p = (x, z)$ is calculated from

$$\Phi_{\text{sim}}(x,z) = \sum_i \Phi_Y + \sum_j \Phi_C + \Phi_{out} \qquad (12)$$

where the sum $i$ runs over all the ions, $\Phi_Y$ is the Yukawa potential given in Eq. 3, the sum $j$ runs over the dust grains which interact through the Coulomb potential $\Phi_C$ given in Eq. 4, and $\Phi_{out}$ is the potential of the uniformly distributed ions outside the simulation region. At low drift velocities, a positive potential region forms between the two dust grains when the particles are far apart. The region then disappears as the particles approach each other (left column). The positive potential region between the two particles becomes less pronounced as the drift velocity increases, and disappears for the highest drift speed $v_{dr}$ = 1.0 M.

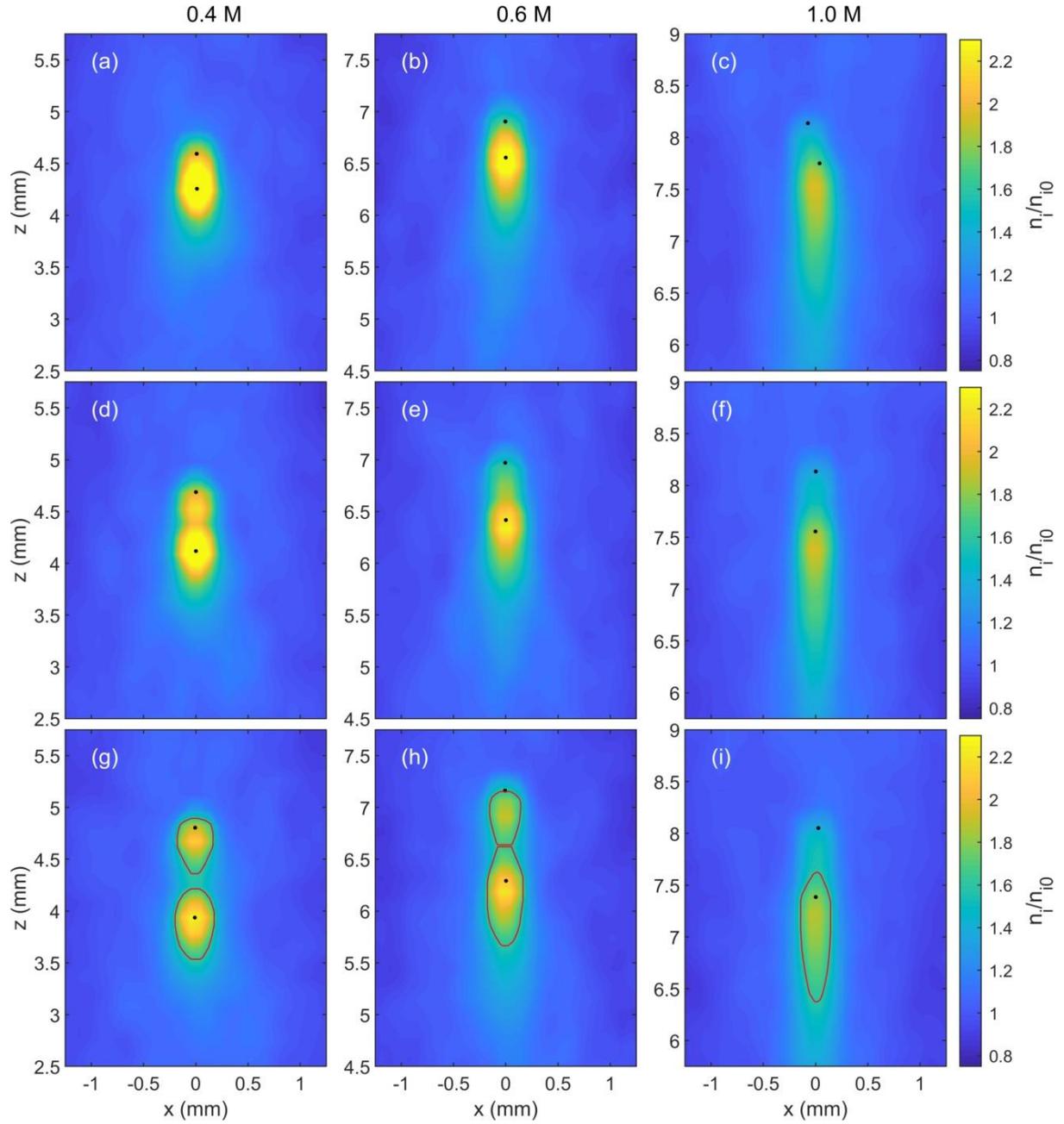

Fig. 5. Ion density maps for two particles: (a-c) vertical particle separation $\Delta z \sim 0.3\, \lambda_{De}$, (d-f) $\Delta z \sim 0.5\, \lambda_{De}$, (g-i) $\Delta z \sim 0.75\, \lambda_{De}$. The ion drift velocity changes from left to right as (left) $v_{dr} = 0.4$ M, (middle) $v_{dr} = 0.6$ M, and (right) $v_{dr} = 1.0$ M. In (g-i), contour lines mark the region of the ion wake where $n_i/n_0 = 1.6$. Videos of the changing ion densities during the simulations are available online. (Multimedia view)

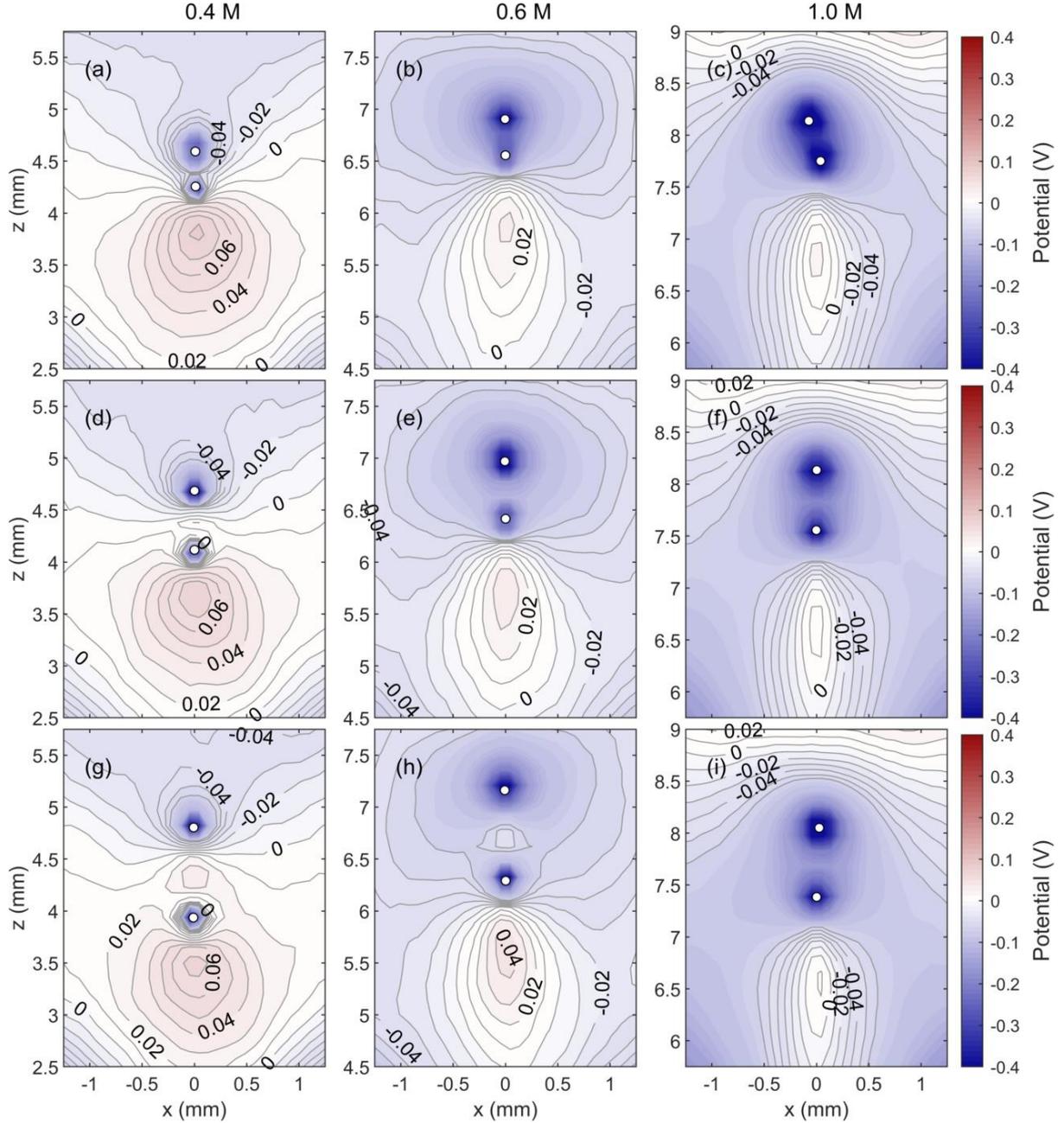

Fig. 6. Electrostatic potential maps for ions and two charged particles: (a-c) vertical particle separation $\Delta z = 0.3 \lambda_{De}$, (d-f) $\Delta z = 0.5 \lambda_{De}$, (g-i) $\Delta z = 0.75 \lambda_D e$. The ion drift velocity changes from left to right as (left) $v_{dr} = 0.4$ M, (middle) $v_{dr} = 0.6$ M, and (right) $v_{dr} = 1.0$ M. The contour lines indicate levels from -0.06:0.01:0.08 V. Videos of the changing potential structure during the simulations are available online. Potential is relative to the average ion potential of 2.8 V. (Multimedia view)

B. Dust Charge

Figure 7 shows the charge of the two particles (upper panels) as well as their relative vertical separation (lower panels) as a function of time. Initially, the particles are vertically aligned and the lower particle oscillates up and down within the wake of the upper particle. The charge on the lower particle is significantly reduced as it approaches the upper particle, while the charge on the upper particle remains relatively constant. Figure 8 shows the charge on the two vertically aligned particles during this time (before the laser push is applied) in units of $Q_0$, the average charge on the upstream particle. The decharging of the downstream particle (shown in blue) is almost linearly proportional to the vertical separation between the two particles. Interestingly, there appears to be a hysteresis in the charge, depending on whether the downstream particle is approaching or receding from the upstream particle. The velocity of the particles is less than a few cm/s, so this effect is not due to a difference in the relative streaming velocity of the ions with respect to the dust. Rather it is likely due to the fact that the lower dust grain is either approaching or receding from the region of high ion density in the wakefield. Note that at the slowest ion drift speed, the charge on the upstream particle is also affected by the location of the downstream particle, with the charge on the upstream particle *increasing* slightly as the particle separation decreases.

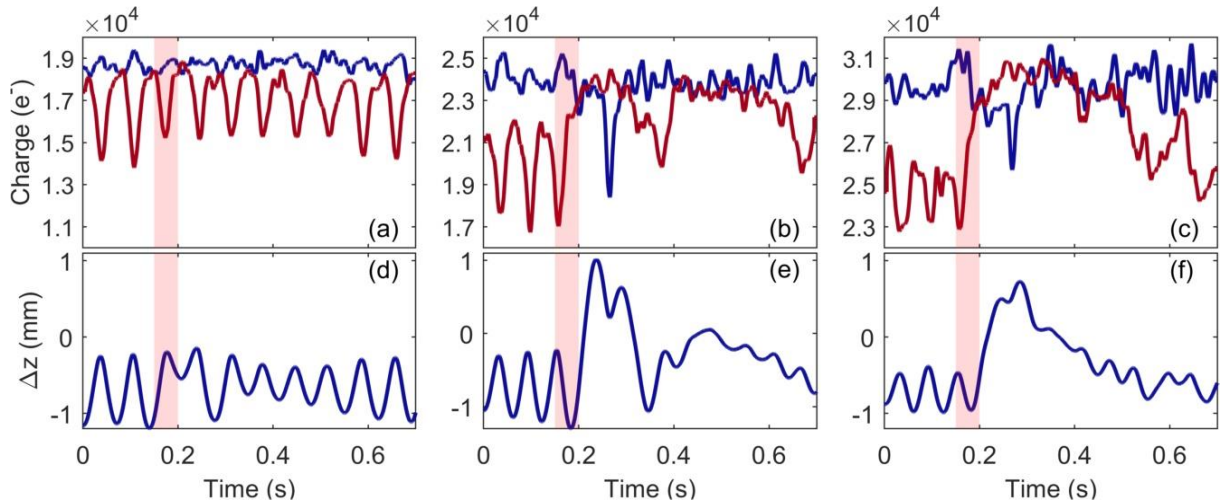

Fig. 7. Charge (top row) and relative vertical separation (bottom row) for two dust particles for three different ion flow velocities (a) $v_{dr} = 0.4$ M, (b) $v_{dr} = 0.6$ M, and (c) $v_{dr} = 1.0$ M. The shaded region indicates the time that the laser acceleration acts on the lower particle. In the upper panels, the charge on the (initial) upper particle (P1) is shown by the blue line, while the red line gives the charge on the (initial) lower particle P2. The relative position $\Delta z = z_2 - z_1$.

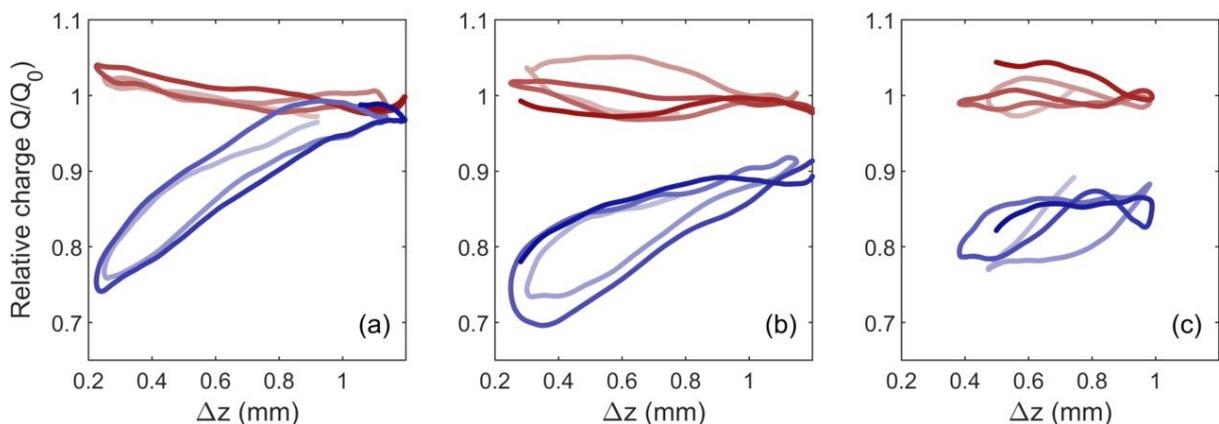

Fig. 8. Charges on vertically aligned particles relative to the average equilibrium charge on the upstream particle for (a) $v_{dr} = 0.4$ M, (b) $v_{dr} = 0.6$ M, and (c) $v_{dr} = 1.0$ M. Time progression is indicated by shade changing from light to dark.

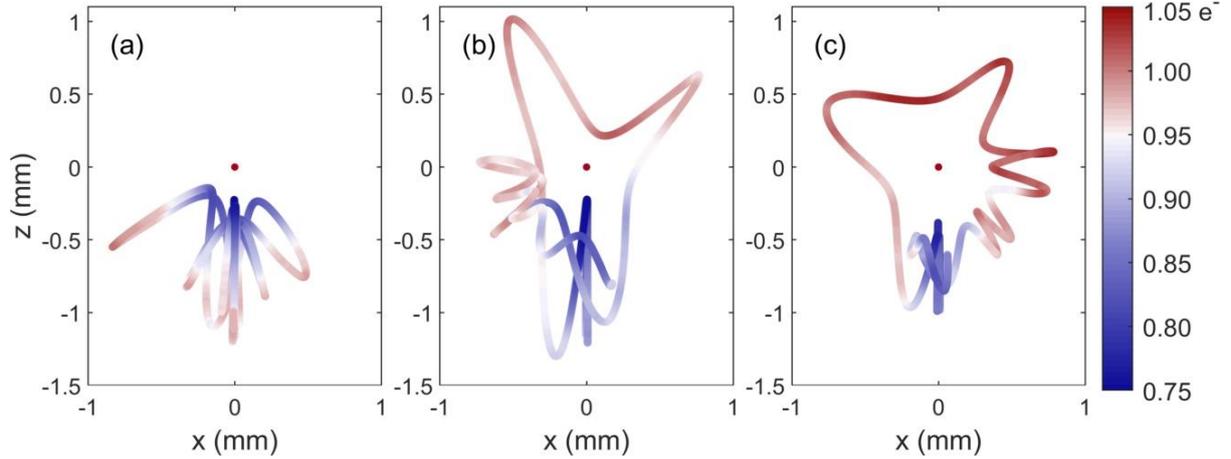

Fig. 9. Relative charge of P2 to $Q_0$, the average equilibrium charge of P1, over the particle trajectory for (a) $v_{dr}$ = 0.4 M, (b) $v_{dr}$ = 0.6 M, and (c) $v_{dr}$ = 1.0 M. Red denotes charge within $\pm 5\%$ of $Q_0$, while blue indicates the region within the ion wakefield where P2 is decharged, $Q_2 < 0.95\, Q_0$. The location of P1 is marked by a red dot.

As the lower particle moves out of upstream particle's wake field, its charge increases and fluctuates about $Q_0$. Figure 9 shows the charge on P2 over the entire trajectory. The region where the ion wake decharges the lower grain is clearly distinguishable, with the horizontal extent of the wakefield region decreasing as the ion drift velocity increases. This is more clearly illustrated in Fig. 10, where the relative charge on the downstream particle is shown as a function of relative horizontal separation $\Delta x$ for several different relative vertical separations, $\Delta z$. The data shown here is for multiple simulation runs, with data taken over the entire trajectory, in which case the particles may change positions. In this case, the charge on the downstream particle is designated $Q_{dn}$ and the charge on the upstream particle is $Q_{up}$. The smallest drift velocity, $v_{dr}$ = 0.4 M, produces a wake that broadens in the horizontal direction as the vertical separation decreases, and the maximum decharging (minimum value of $Q_{dn}/Q_{up}$) occurs at the smallest particle separation (Fig. 10a). At the highest drift velocity, $v_{dr}$ = 1.0 M (Fig. 11c), the horizontal extent of the wake is nearly constant with vertical separation $\Delta z$, and the charge ratio $Q_{dn}/Q_{up}$ reaches its minimum value at approximately $\Delta z = 0.5\, \lambda_{De}$.

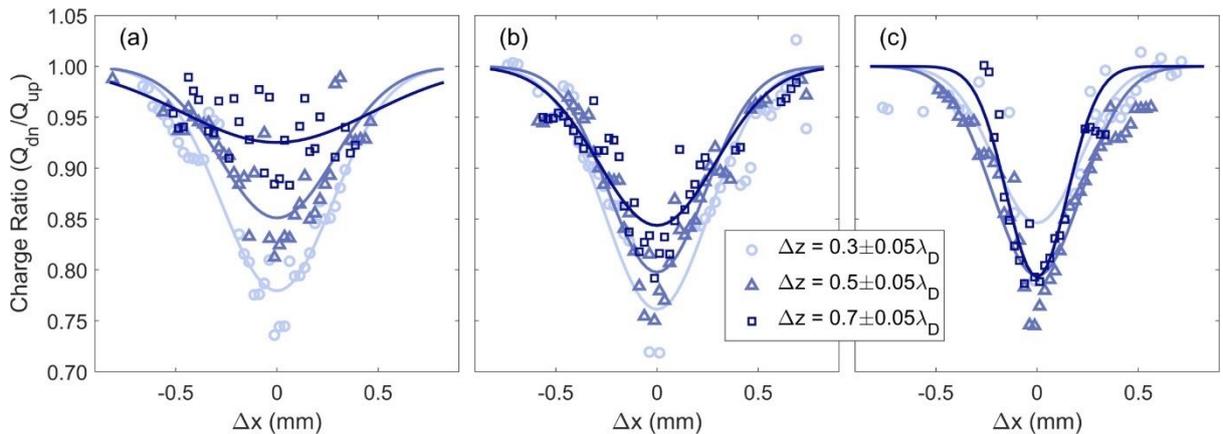

Fig. 10. Relative charge on lower particle as it moves through the wake of the upper particle. The ion drift velocity is (a) $v_{dr}$ = 0.4 M, (b) $v_{dr}$ = 0.6 M, and (c) $v_{dr}$ = 1.0 M. Increasing vertical separation between the particles is indicated by darker shades. The data points represent the average minimum charge at a given horizontal separation, while the lines are a normal fit to the data.

C.     Characteristics of the Ion Wake

As two vertically-aligned particles approach each other in the z-direction, the size of the wake changes and the location of the maximum ion density shifts, even forming downstream of the second particle. As shown in Fig. 11 (a-c), at large particle separations the maximum ion density is at a location $\Delta L = 0.1\text{-}0.2\ \lambda_{De}$, with $\Delta L$ increasing as the drift velocity increases. When the particle separation decreases below $\sim 0.4\ \lambda_{De}$, the ion focusing region makes a sudden transition to a point at or below the location of the second particle, indicated by the points lying above the dashed line. The maximum ion density in the ion wakes of the two particles is shown in Fig. 11 (d-f). At low drift velocity (Fig. 11d), the two particles have similar maxima at large distances. As the two particles approach each other, the ion density beneath P2 grows in magnitude, until a single wake is formed for interparticle separations $< 0.4\ \lambda_{De}$. At higher drift velocity, $v_{dr} = 0.6, 1.0$ M, the maximum ion density beneath P1 is always less than that of P2, and relatively constant, while the magnitude of the wake downstream of P2 grows with decreasing particle separation (Fig. 11 e,f).

An estimate of the excess positive charge in the ion cloud associated with each grain is calculated from the charge contained within a region downstream of a dust grain where the ion density $n_i > 1.6\ n_0$, as indicated by the contour lines in Fig. 5 g-i. The radial extent of this region $r_w$ (assumed to be azimuthally symmetric) and axial extent are shown in Fig. 12. The radial extent of the wake tends to be fairly constant with particle separation (Fig. 12 a-c). However, as the two particles approach each other, the vertical extent of the wake downstream of P1 is reduced, while the vertical extent of the wake downstream of P2 increases slightly. The wake below P2 becomes more elongated with ion drift velocity, increasing in vertical extent. Figure 13 shows the total excess charge in the wake downstream of each particle,

$$q_w = 2\pi e \sum_j (n_{i,j} - n_0)\, r_j \Delta r \Delta z \qquad (13)$$

(where $j$ is the index of the grid points with $n_i > 1.6\ n_o$ and $\Delta r$ and $\Delta z$ are the dimensions of each grid cell) normalized by the charge on the dust grain. As the separation between the particles decreases, the excess charge below P1 decreases slightly (even though the maximum charge density remains relatively constant, as shown in Fig. 11 d-f), while the excess charge below P2 increases. Below a distance of $\sim 0.4\lambda_{De}$, a single wake is created, with a charge that generally exceeds the wake charge below P2 alone. The limiting value of the total charge contained within the wake downstream of P1 and P2 at $\Delta z = \lambda_{De}$ is given in Table I.

Table I. Positive charge contained within the ion wake downstream of each particle for particle separation $\Delta z = \lambda_{De}$, for given ion flow speeds.

| $v_{dr}$ | 0.4 M | 0.6 M | 1.0 M |
|---|---|---|---|
| $q_{w1}/Q_{d1}$ | 0.32 | 0.12 | 0.043 |
| $q_{w2}/Q_{d2}$ | 0.33 | 0.30 | 0.32 |

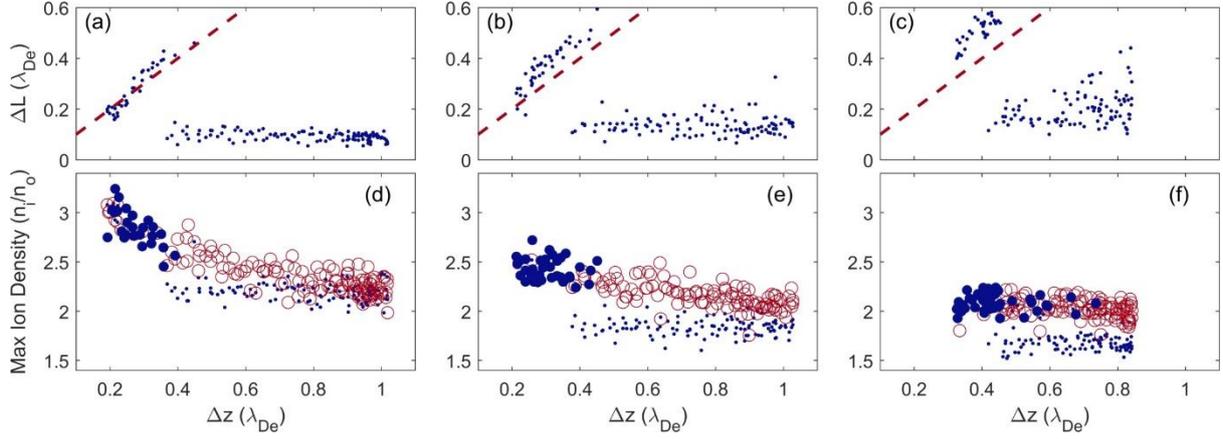

Fig. 11. (a-c) Location of the maximum ion density beneath the upper particle as a function of interparticle separation. Points lying above the dashed line (with slope = 1) indicate positions where there is a single ion focusing region below the downstream particle. (d-f) Maximum ion density in the wake beneath P1 (blue points) and P2 (red open circles). Distances where there is only a single maximum in the ion wake are indicated by a filled circle. The ion drift velocity is (a,d) $v_{dr} = 0.4$ M, (b,e) $v_{dr} = 0.6$ M, and (c,f) $v_{dr} = 1.0$ M.

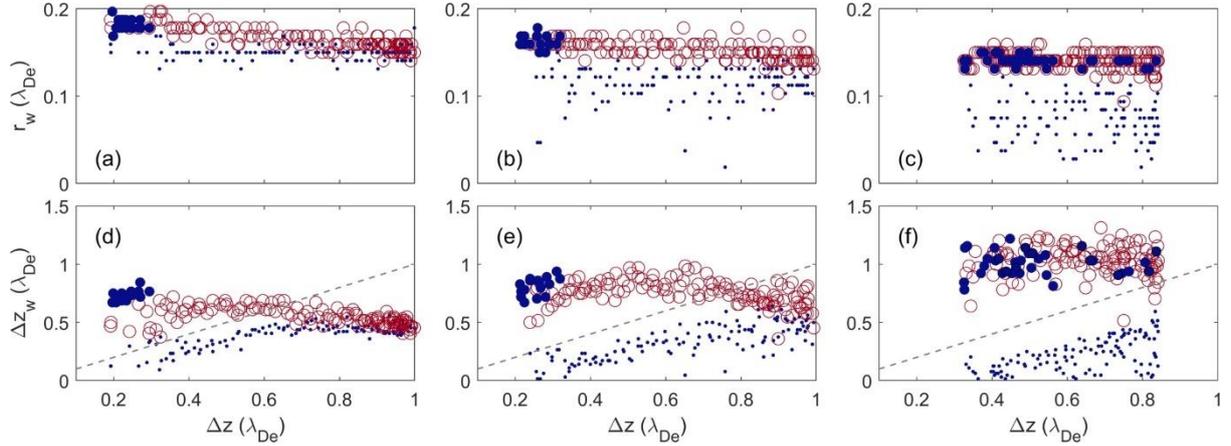

Fig. 12. (a,b,c) Radial and (d,e,f) vertical extent of the wake, defined by the region where the ion density > 1.6 $n_0$ (see Fig. 6 g-i). The ion drift velocity is (a,d) $v_{dr} = 0.4$ M, (b,e) $v_{dr} = 0.6$ M, and (c,f) $v_{dr} = 1.0$ M. The wake beneath P1 is marked by (blue) points and wake beneath P2 are marked by (red) open circles. Filled circles indicate points when there is only a single maximum in the ion wake. The dashed line indicates the distance between the two particles.

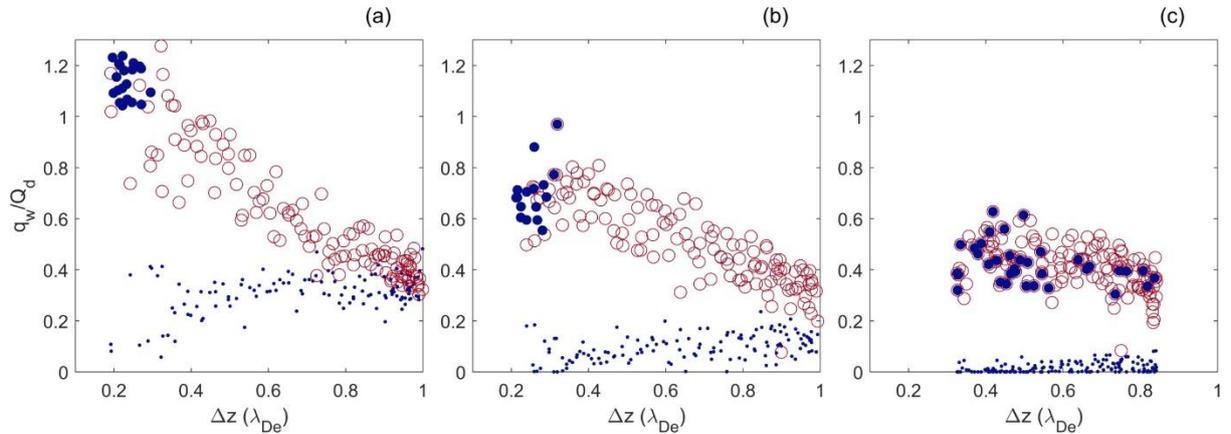

Fig. 13. Excess positive charge contained within the wake beneath P1 (blue points) and P2 (red open circles). At close distances there is only a single ion wake, indicated by filled blue circles. The ion drift velocity is (a) $v_{dr} = 0.4$ M, (b) $v_{dr} = 0.6$ M, and (c) $v_{dr} = 1.0$ M.

## D. Accuracy of the Point Charge Model

The changing location and magnitude of the focused ion wake has implications for the particle dynamics. Many numerical models of coupled dust motion make use of the "point-charge model", where the positive ion cloud downstream of a dust particle is assumed to have a fixed charge $q_0$ and fixed location $\Delta L$ beneath each particle [41]–[44]. This may be a fair approximation if particles are in a horizontal plane. As shown in Fig. 14 abc, the ion densities downstream of two particles with approximate horizontal alignment (small $\Delta z$) are nearly the same, though for the small horizontal spacing of $0.4 - 0.5\ \lambda_{De}$ shown here, there is a single combined positive potential region downstream of the two grains (Fig 14 def).

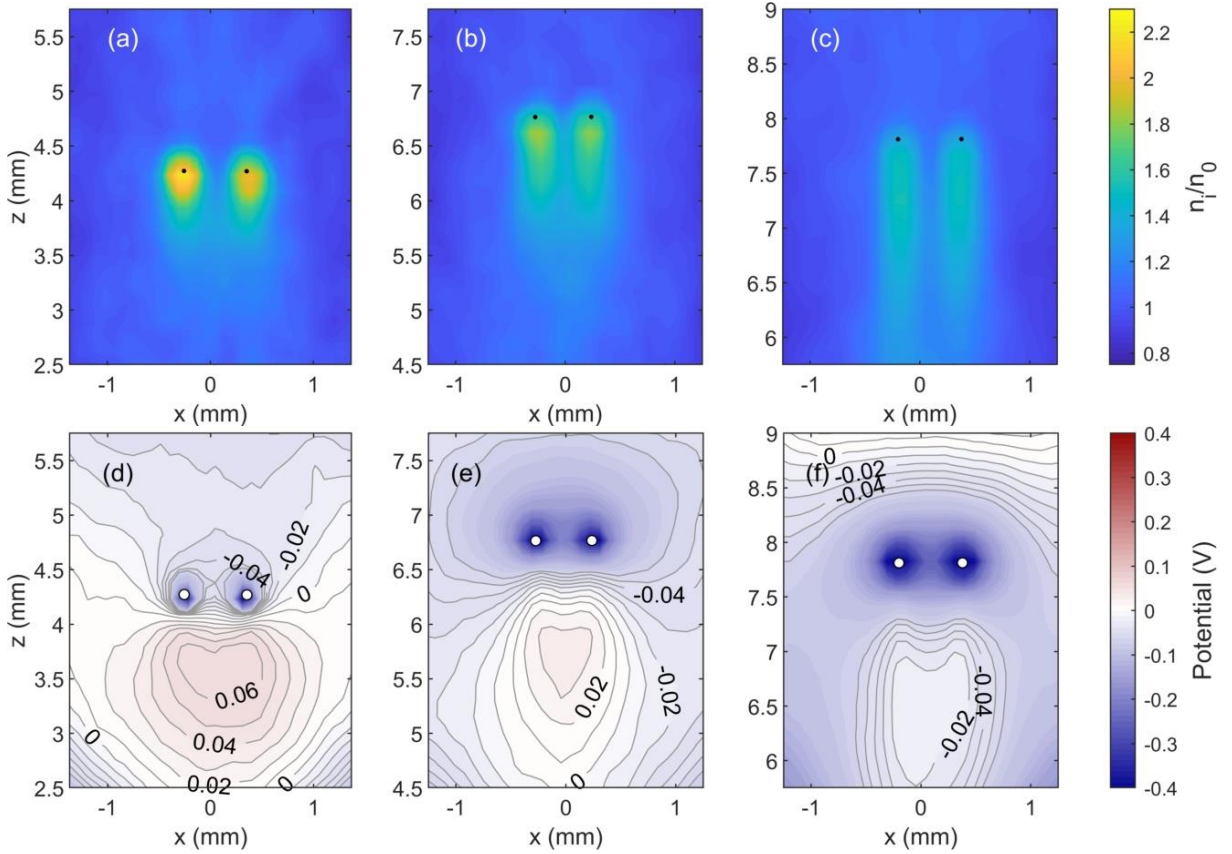

Fig. 14. (a,b,c) Ion density and (d,e,f) potential maps for horizontally-aligned particles. The potential is measured relative to the average ion potential of 2.8 V. Contour lines designate level -0.06:0.01:0.08 V. The ion drift velocity and horizontal displacement are (a,d) $v_{dr}$ = 0.4 M, $\Delta x = 0.5\ \lambda_{De}$ (b,e) $v_{dr}$ = 0.6 M, $\Delta x = 0.4\ \lambda_{De}$ and (c,f) $v_{dr}$ = 1.0 M, $\Delta x = 0.5\ \lambda_{De}$.

For vertically aligned particles, the result is more complicated as the location and magnitude of the wake charge are affected by the downstream particle. Here we compare our simulation results, $\Phi_{sim}$ (calculated from Eq. 12 on the symmetry axis connecting the two particles), to (1) a Coulomb potential, (2) a spherical point charge model, and (3) an ellipsoid point charge model. The Coulomb potential considered for case (1) is given by

$$\Phi_C(z) = \sum_{j=1,2} \frac{1}{4\pi\epsilon_0} \frac{Q_j}{|z-z_j|} \qquad (14)$$

where $j$ is the index of the two dust particles. For case (2) the positive point charge is represented by a sphere of charge $q_w$ and radius $r_w$, as shown in Figs. 12 and 13, centered at the maximum in the ion density, as shown in Fig. 11. The resulting potential as a function of position $z$ along the symmetry axis is given by

$$\Phi_{pt} = \Phi_Y + \sum_{j=1,2} \begin{cases} \frac{1}{4\pi\epsilon_0} \frac{q_{w,j}}{|z-z_{w,j}|}, & z > r_{w,j} \\ \frac{1}{4\pi\epsilon_0} \frac{Q_{w,j}}{2r_{w,j}} \left(3 - \frac{(z-z_{w,j})^2}{r_{w,j}^2}\right), & z \leq r_{w,j} \end{cases}. \quad (16)$$

In case (3) the positive point charge is represented by an ellipsoid of charge $q_w$, with semimajor axes $a = b = r_w$ and $c = \Delta z_w/2$, centered at the maximum in the ion density. The potential on the symmetry axis as a function of $z$ is given by

$$\Phi_{el} = \Phi_Y + \sum_{j=1,2} q_w \int_{0,\lambda}^{\infty} \left[1 - \frac{(z-z_{w,j})^2}{c^2+s}\right] \frac{ds}{\sqrt{\varphi(s)}} \quad (17)$$

where $\varphi(s) = (a^2 + s)(b^2 + s)(c^2 + s)$, and the lower integration bound is 0 if a point is inside the ellipse or $\lambda$ if a point is outside the ellipse; $\lambda$ is the greatest root of $f(s) = \frac{z^2}{c^2+s} - 1 = 0$ [45]. The simulation results for the potential upstream and downstream of the upper particle, $\Phi_{sim}$ (Eq. 12), are shown in Fig. 15 by the solid blue lines, for varying particle separations. Within the simulation region, the potential from the ions is smaller at the bottom of the cylinder than at the top (especially for the higher drift velocities), with a slope that ranges from -3.5 V/m for $v_d = 0.4\ M$ to 35 V/m for $v_d = 1.0\ M$. This slope is subtracted from the total potential, for comparison with the Yukawa and point charge potentials which assume a uniform background potential. For comparison purposes, the zero of each potential is set by its value at a distance $z = \lambda_{De}$, and then normalized by $V_o = -1/(4\pi\epsilon_o)\, (Q_{up}/\lambda_{De})$, the magnitude of the Coulomb potential of the upper particle at a distance of $z = \lambda_{De}$.

At the greatest ion flow speed $v_d = 1.0$ M (Fig. 15 c), the potential downstream of P1 is well-represented by the point charge model, as all of the excess ion density tends to be downstream of P1. The spherical point charge model is less accurate as the ion drift speed decreases, as it does not fully capture the increased ion density which extends *upstream* of each particle. In this case, the ellipsoidal point charge model yields the best match at small interparticle spacing and does a better job of capturing the overall shape of the potential for larger particle separations. Upstream of P1, the ion shielding becomes less effective as the drift speed of the ions increases (Figs. 15 b,c), and the upstream potential is matched fairly well by the Coulomb potential.

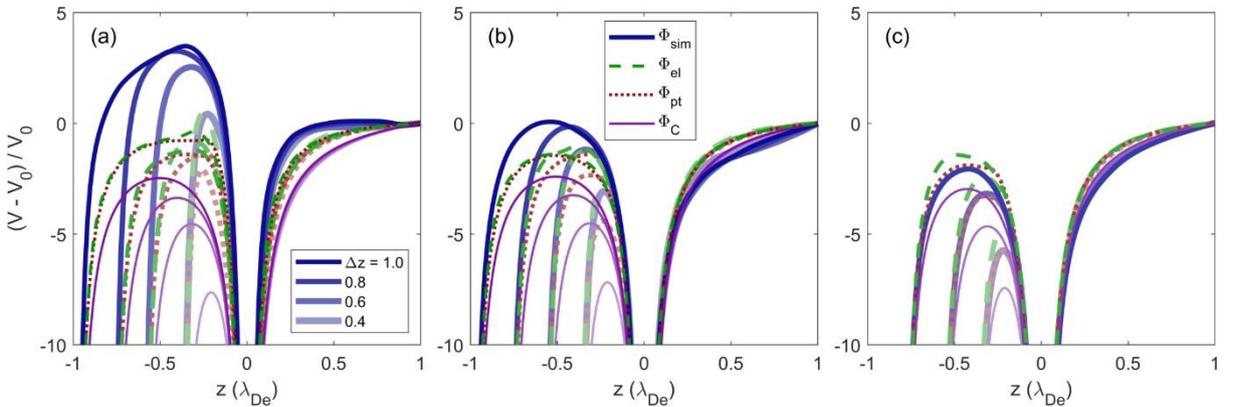

Fig. 15. Total potential calculated from the simulation results (thick solid blue lines) upstream and downstream of the upper particle (located at $z = 0$) for varying particle separations. The point charge potential using ellipsoidal positive region is indicated by green dashed lines, the point charge potential using a spherical positive region is indicated by red dotted lines, and the combined Coulomb potential for the two particles is indicated by the thin purple lines. For clarity, the potential downstream of the second particle is not shown. The ion drift velocity is (a) $v_{dr} = 0.4$ M, (b) $v_{dr} = 0.6$ M, and (c) $v_{dr} = 1.0$ M.

E.   Effects on Dust Dynamics

The ion wake field has long been successfully employed to explain the non-reciprocal interaction between upstream and downstream particles, as well as the attractive horizontal force that the upstream particle exerts on the downstream particle [41], [46]–[49]. The extent of this effect can be examined by calculating the net electric force exerted by the ions on a dust grain from the gradient of the ion potential at the location of the particle, $\boldsymbol{F}_{id} = Q_d \nabla \Phi_i$.

Figure 16 shows the horizontal acceleration calculated for the grains as a function of their relative position for the three different ion flow speeds. The maps were generated by overlaying the region where the data was collected with a grid with spacing of 100 μm. The average acceleration $(F_{dd,x} + F_{id,x})/m_d$ was calculated for all data points within a radius of 70.7 μm of each grid point. The effect of the ion wake field downstream of the particle (marked by the black dot at the origin) is readily apparent in Fig. 16, where positive values indicate acceleration towards the right. There is a wedge-shaped region below the upstream grain over which the attractive force from the wake potential exceeds the repulsive particle interaction force. The location of this region moves downward and becomes narrower as the ion flow speed increases.

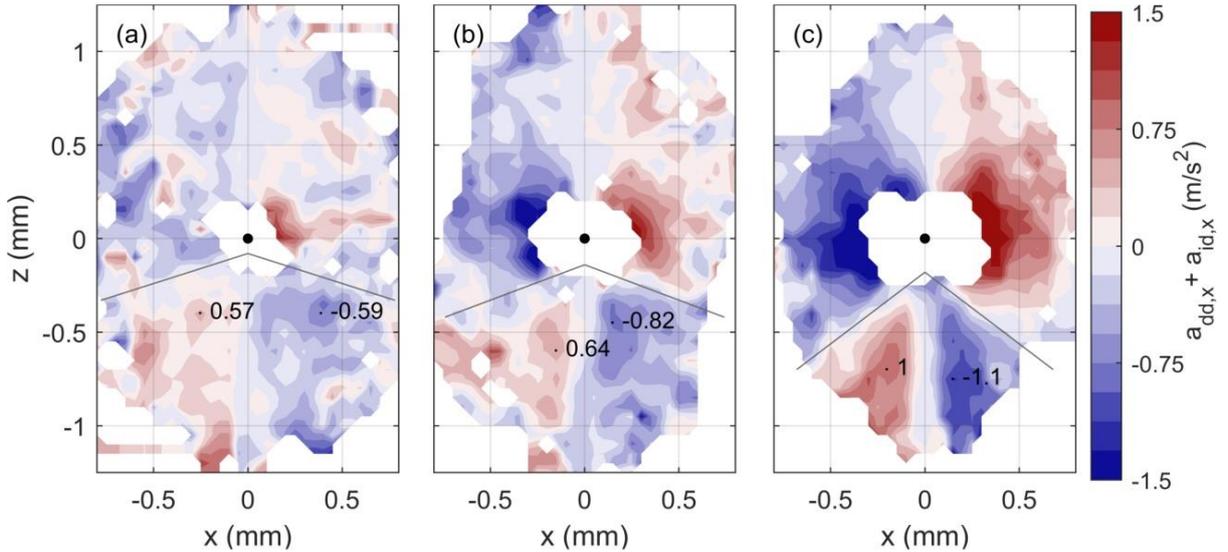

Fig. 16. Map of horizontal acceleration of a particle for positions relative to a second particle, marked by the black circle. Positive values (red) indicate acceleration towards the right, while negative values (blue) indicate acceleration towards the left. The ion drift velocity is (a) $v_{dr} = 0.4$ M, (b) $v_{dr} = 0.6$ M, and (c) $v_{dr} = 1.0$ M. The lines indicate the boundary of the region over which the ion wake exerts an attractive force on the downstream grain.

In the vertical direction, the electric force exerted by the ions on the dust grains is always downwards, reflecting the positive potential region formed downstream of the dust. However, there is an asymmetry in the drag which is exerted on each particle, as shown in Fig. 17, which is in the center-of-mass (COM) system. Since the particles have the same mass, in the COM system $(x_1, z_1) = -(x_2, z_2)$. An upstream particle ($z_{COM} > 0$) always experiences a larger force from the ions than a downstream particle. As two particles approach each other in the vertical direction, the downward force on the upstream particle increases, as does the upward force on the downstream particle. The net effect of the ion-dust force in the vertical direction is to push the two particles together, in effect reducing the repulsion between the grains. This asymmetry lessens as the ion drift speed increases.

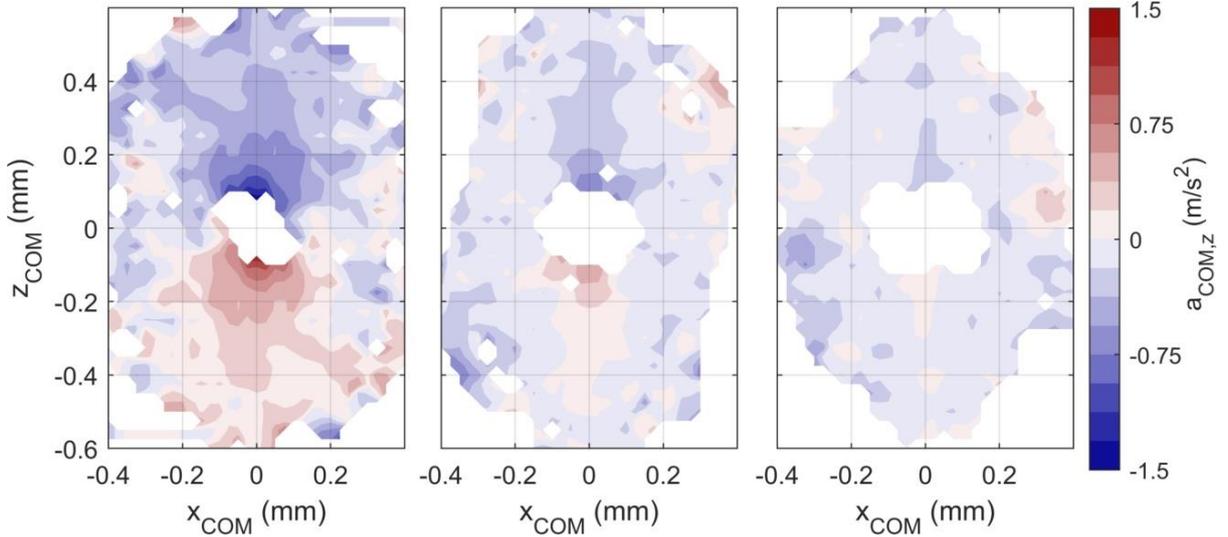

Fig. 17. Map of the vertical acceleration due to the ion force in the center-of-mass system. Positive values (red) indicate upward acceleration, while negative values (blue) indicate downward acceleration. The ion drift velocity is (a,d) $v_{dr} = 0.4$ M, (b,e) $v_{dr} = 0.6$ M, and (c,f) $v_{dr} = 1.0$ M.

IV. Conclusion

A numerical model is presented which simultaneously tracks the dynamics of ions and dust in a plasma environment with flowing ions. The dust charge and the ion wakefield are self-consistently calculated from the ion dynamics, allowing for detailed analysis of the wakefield-mediated interaction as the structural configuration of the dust grains changes, as illustrated in Figs. 5 and 6.

The charging and decharging of the downstream grain is tracked over the entire dust trajectory, as illustrated in Figs. 7 and 9, and is mapped as a function of vertical and horizontal separation, as shown in Figs. 8 and 10. At subsonic drift speeds, the horizontal extent over which decharging occurs, and the magnitude of the decharging, increases as the particles approach each other (Fig. 10). As the ion drift speed approaches $v_{dr} = 1.0$ M, the magnitude of the decharging and the horizontal extent over which it occurs are relatively constant. An interesting feature is the apparent hysteresis in the charging/decharging of the lower grain, especially evident at subsonic ion flow speeds, as shown in Fig. 8.

The location and magnitude of the enhanced ion density changes with ion flow velocity as well as relative dust position (Fig. 11). For large particle separations, $\Delta z \geq \lambda_{De}$, the radial and axial extent, location, and maximum ion density of the wake all approach a constant value, as seen in Figs. 11 and 12. For the downstream grain, these values increase as the particle separation decreases. As the wake expands, less charge is contained in the wake of the upstream particle, $q_{w1}$, while the charge in the wake of the downstream particle, $q_{w2}$, grows (Fig. 13). For subsonic ion flow and close particle separations, the ion focusing regions downstream of each of the particles merge, creating a single maximum positive potential downstream of the second particle. For ion flow with $v_D = 1.0$ M, however, the extent and maximum density of the wake is relatively independent of the particle separation, with the wake almost entirely downstream of P2. The positive space charge region of the ion wake can be successfully modeled by a spherical point charge model for ion drift speeds $v_d \geq 1.0$ M (see Fig. 15). At subsonic flow velocities, the enhanced ion density is better captured by an ellipsoid of positive charge, with best agreement for particle separations $\Delta z < 0.4 \, \lambda_{De}$.

The non-reciprocal interaction force between the two dust grains is readily apparent in both the horizontal (Fig. 16) and vertical (Fig. 17) directions. In the horizontal direction, the downstream grain experiences an attractive force which increases in magnitude as the ion drift speed increases. In the vertical direction, the asymmetry in the interaction force is due to the difference in the force exerted by the ions on the dust. The upstream particle always experiences a larger downward force, while the force is reduced for the downstream particle. The effect of this asymmetry in the vertical force decreases with the ion drift speed.

The simulated conditions shown here were for two dust grains confined within a glass box placed on the lower electrode of a rf cell. However, the boundary conditions in the numerical model can easily be adapted for different experiments such as the DC glow discharge in the PK-4 experiment onboard the International Space Station [50]. In this experiment, the polarity of the DC electrodes can be rapidly switched in order to trap the dust grains in the field of view of the cameras. This can create an ion wake both upstream and downstream of the dust grains, resulting in a homogeneous-to-string structural transition of the dust cloud, an interesting condition for further research.

Acknowledgements

Thanks to Peter Hartmann and Marlene Rosenberg for useful discussion about this work. Support from NSF Grant numbers 1707215 and 174023 and NASA Grant number 1571701 is gratefully acknowledged.

Bibliography

[1] P. Hartmann *et al.*, "Crystallization Dynamics of a Single Layer Complex Plasma," *Phys. Rev. Lett.*, vol. 105, p. 115004, 2010.
[2] J. H. Chu and L. I, "Direct observation of Coulomb crystals and liquids in strongly coupled rf dusty plasmas," *Phys. Rev. Lett.*, vol. 72, no. 25, pp. 4009–4012, Jun. 1994.
[3] A. P. Nefedov, O. F. Petrov, V. I. Molotkov, and V. E. Fortov, "Formation of liquidlike and crystalline structures in dusty plasmas," *J. Exp. Theor. Phys. Lett.*, vol. 72, no. 4, pp. 218–226, Aug. 2000.
[4] J. Kong, T. W. Hyde, L. Matthews, K. Qiao, Z. Zhang, and A. Douglass, "One-dimensional vertical dust strings in a glass box," *Phys. Rev. E*, vol. 84, no. 1, p. 016411, Jul. 2011.
[5] H. Thomas, G. E. Morfill, V. Demmel, J. Goree, B. Feuerbacher, and D. Möhlmann, "Plasma Crystal: Coulomb Crystallization in a Dusty Plasma," *Phys. Rev. Lett.*, vol. 73, no. 5, pp. 652–655, Aug. 1994.
[6] G. A. Hebner and M. E. Riley, "Measurement of attractive interactions produced by the ion wakefield in dusty plasmas using a constrained collision geometry," *Phys. Rev. E*, vol. 68, no. 4, p. 046401, Oct. 2003.
[7] A. Melzer, V. A. Schweigert, and A. Piel, "Measurement of the Wakefield Attraction for 'Dust Plasma Molecules,'" *Phys. Scr.*, vol. 61, no. 4, pp. 494–501, Apr. 2000.

[8] M. Kroll, J. Schablinski, D. Block, and A. Piel, "On the influence of wakefields on three-dimensional particle arrangements," *Phys. Plasmas*, vol. 17, no. 1, p. 013702, Jan. 2010.

[9] A. Melzer, A. Schella, and M. Mulsow, "Nonequilibrium finite dust clusters: Connecting normal modes and wakefields," *Phys. Rev. E*, vol. 89, no. 1, p. 013109, Jan. 2014.

[10] K. Qiao, J. Kong, E. V. Oeveren, L. S. Matthews, and T. W. Hyde, "Mode couplings and resonance instabilities in dust clusters," *Phys. Rev. E*, vol. 88, no. 4, p. 043103, Oct. 2013.

[11] K. Qiao, J. Kong, Z. Zhang, L. S. Matthews, and T. W. Hyde, "Mode Couplings and Conversions for Horizontal Dust Particle Pairs in Complex Plasmas," *IEEE Trans. Plasma Sci.*, vol. 41, no. 4, pp. 745–753, Apr. 2013.

[12] F. Melandsø, "Heating and phase transitions of dust-plasma crystals in a flowing plasma," *Phys. Rev. E*, vol. 55, no. 6, pp. 7495–7506, Jun. 1997.

[13] F. Melandsø and J. Goree, "Polarized supersonic plasma flow simulation for charged bodies such as dust particles and spacecraft," *Phys. Rev. E*, vol. 52, no. 5, pp. 5312–5326, Nov. 1995.

[14] V. A. Schweigert, I. V. Schweigert, A. Melzer, A. Homann, and A. Piel, "Alignment and instability of dust crystals in plasmas," *Phys. Rev. E*, vol. 54, no. 4, pp. 4155–4166, Oct. 1996.

[15] I. H. Hutchinson, "Intergrain forces in low-Mach-number plasma wakes," *Phys. Rev. E*, vol. 85, no. 6, p. 066409, Jun. 2012.

[16] V. R. Ikkurthi, K. Matyash, A. Melzer, and R. Schneider, "Computation of charge and ion drag force on multiple static spherical dust grains immersed in rf discharges," *Phys. Plasmas*, vol. 17, no. 10, p. 103712, Oct. 2010.

[17] W. J. Miloch, M. Kroll, and D. Block, "Charging and dynamics of a dust grain in the wake of another grain in flowing plasmas," *Phys. Plasmas*, vol. 17, no. 10, p. 103703, Oct. 2010.

[18] I. H. Hutchinson and C. B. Haakonsen, "Collisional effects on nonlinear ion drag force for small grains," *Phys. Plasmas*, vol. 20, no. 8, p. 083701, Aug. 2013.

[19] I. H. Hutchinson, "Forces on a Small Grain in the Nonlinear Plasma Wake of Another," *Phys. Rev. Lett.*, vol. 107, no. 9, p. 095001, Aug. 2011.

[20] W. J. Miloch, M. Kroll, and D. Block, "Charging and dynamics of a dust grain in the wake of another grain in flowing plasmas," *Phys. Plasmas*, vol. 17, no. 10, p. 103703, Oct. 2010.

[21] W. J. Miloch and D. Block, "Dust grain charging in a wake of other grains," *Phys. Plasmas*, vol. 19, no. 12, p. 123703, Dec. 2012.

[22] W. J. Miloch, "Simulations of Several Finite-sized Objects in Plasma," *Procedia Comput. Sci.*, vol. 51, pp. 1282–1291, Jan. 2015.

[23] W. J. Miloch, H. Jung, D. Darian, F. Greiner, M. Mortensen, and A. Piel, "Dynamic ion shadows behind finite-sized objects in collisionless magnetized plasma flows," *New J. Phys.*, vol. 20, no. 7, p. 073027, Jul. 2018.

[24] A. Piel, "Molecular dynamics simulation of ion flows around microparticles," *Phys. Plasmas*, vol. 24, no. 3, p. 033712, Mar. 2017.

[25] A. Piel, F. Greiner, H. Jung, and W. J. Miloch, "Molecular dynamics simulations of wake structures behind a microparticle in a magnetized ion flow. I. Collisionless limit with cold ion beam," *Phys. Plasmas*, vol. 25, no. 8, p. 083702, Aug. 2018.

[26] A. Piel, H. Jung, and F. Greiner, "Molecular dynamics simulations of wake structures behind a microparticle in a magnetized ion flow. II. Effects of velocity spread and ion collisions," *Phys. Plasmas*, vol. 25, no. 8, p. 083703, Aug. 2018.

[27] I. H. Hutchinson, "Intergrain forces in low-Mach-number plasma wakes," *Phys. Rev. E*, vol. 85, no. 6, p. 066409, Jun. 2012.
[28] I. H. Hutchinson, "Ion collection by a sphere in a flowing plasma: I. Quasineutral," *Plasma Phys. Control. Fusion*, vol. 44, no. 9, p. 1953, 2002.
[29] A. Piel and J. A. Goree, "Collisional and collisionless expansion of Yukawa balls," *Phys. Rev. E*, vol. 88, no. 6, p. 063103, Dec. 2013.
[30] S. Ratynskaia *et al.*, "Experimental Determination of Dust-Particle Charge in a Discharge Plasma at Elevated Pressures," *Phys. Rev. Lett.*, vol. 93, no. 8, Aug. 2004.
[31] S. A. Khrapak *et al.*, "Particle charge in the bulk of gas discharges," *Phys. Rev. E*, vol. 72, no. 1, p. 016406, Jul. 2005.
[32] M. Gatti and U. Kortshagen, "Analytical model of particle charging in plasmas over a wide range of collisionality," *Phys. Rev. E*, vol. 78, no. 4, Oct. 2008.
[33] Z. Donkó, "Particle simulation methods for studies of low-pressure plasma sources," *Plasma Sources Sci. Technol.*, vol. 20, no. 2, p. 024001, 2011.
[34] A. V. Phelps, "The application of scattering cross sections to ion flux models in discharge sheaths," *J. Appl. Phys.*, vol. 76, no. 2, pp. 747–753, Jul. 1994.
[35] J. V. Jovanović, S. B. Vrhovac, and Z. Lj. Petrović, "Momentum transfer theory of ion transport under the influence of resonant charge transfer collisions: the case of argon and neon ions in parent gases," *Eur. Phys. J. - At. Mol. Opt. Plasma Phys.*, vol. 21, no. 3, pp. 335–342, Dec. 2002.
[36] A. Douglass, V. Land, K. Qiao, L. Matthews, and T. Hyde, "Determination of the levitation limits of dust particles within the sheath in complex plasma experiments," *Phys. Plasmas*, vol. 19, no. 1, pp. 013707-013707–8, Jan. 2012.
[37] P. Hartmann, A. Z. Kovács, J. C. Reyes, L. S. Matthews, and T. W. Hyde, "Dust as probe for horizontal field distribution in low pressure gas discharges," *Plasma Sources Sci. Technol.*, vol. 23, no. 4, p. 045008, Aug. 2014.
[38] J. E. Allen, "Probe theory - the orbital motion approach," *Phys. Scr.*, vol. 45, no. 5, p. 497, May 1992.
[39] T. Matsoukas and M. Russell, "Fokker-Planck description of particle charging in ionized gases," *Phys. Rev. E*, vol. 55, no. 1, pp. 991–994, Jan. 1997.
[40] M. Chen, M. Dropmann, B. Zhang, L. S. Matthews, and T. W. Hyde, "Ion-wake Field inside a Glass Box," *Phys. Rev. E*, vol. 94, no. 3, Sep. 2016.
[41] V. Schweigert, I. Schweigert, A. Melzer, A. Homann, and A. Piel, "Alignment and instability of dust crystals in plasmas," *Phys. Rev. E*, vol. 54, no. 4, pp. 4155–4166, Oct. 1996.
[42] A. V. Ivlev and G. Morfill, "Anisotropic dust lattice modes," *Phys. Rev. E*, vol. 63, no. 1, p. 016409, Dec. 2000.
[43] K. Qiao, J. Kong, E. V. Oeveren, L. S. Matthews, and T. W. Hyde, "Mode couplings and resonance instabilities in dust clusters," *Phys. Rev. E*, vol. 88, no. 4, p. 043103, Oct. 2013.
[44] K. Qiao, J. Kong, J. Carmona-Reyes, L. S. Matthews, and T. W. Hyde, "Mode coupling and resonance instabilities in quasi-two-dimensional dust clusters in complex plasmas," *Phys. Rev. E*, vol. 90, no. 3, p. 033109, Sep. 2014.
[45] W. Cai, "Potential Field of a Uniformly Charged Ellipsoid," p. 18.
[46] G. A. Hebner and M. E. Riley, "Measurement of attractive interactions produced by the ion wakefield in dusty plasmas using a constrained collision geometry," *Phys. Rev. E*, vol. 68, no. 4, p. 046401, Oct. 2003.
[47] V. Steinberg, R. Sütterlin, A. V. Ivlev, and G. Morfill, "Vertical Pairing of Identical Particles Suspended in the Plasma Sheath," *Phys. Rev. Lett.*, vol. 86, no. 20, pp. 4540–4543, May 2001.

[48] K. Qiao, J. Kong, Z. Zhang, L. S. Matthews, and T. W. Hyde, "Mode Couplings and Conversions for Horizontal Dust Particle Pairs in Complex Plasmas," *IEEE Trans. Plasma Sci.*, vol. 41, no. 4, pp. 745–753, 2013.

[49] K. Qiao, J. Kong, L. S. Matthews, and T. W. Hyde, "Mode couplings and resonance instabilities in finite dust chains," *Phys. Rev. E*, vol. 91, no. 5, p. 053101, May 2015.

[50] M. Y. Pustylnik *et al.*, "Plasmakristall-4: New complex (dusty) plasma laboratory on board the International Space Station," *Rev. Sci. Instrum.*, vol. 87, no. 9, p. 093505, Sep. 2016.